\documentclass[useAMS]{mn2e}
\usepackage{graphicx}
\usepackage{times}
\usepackage{amssymb}

\def\gsim { \lower .75ex \hbox{$\sim$} \llap{\raise .27ex \hbox{$>$}} }
\def\lsim { \lower .75ex \hbox{$\sim$} \llap{\raise .27ex \hbox{$<$}} }

\begin{document}

\title[Luminous Halos of Galaxies]{Stars beyond Galaxies: The Origin of
Extended Luminous Halos around Galaxies}

\author[Abadi, Navarro \& Steinmetz]{Mario G. Abadi, $^{1}$\thanks{CITA National
Fellow, on leave from Observatorio Astron\'omico de C\'ordoba and CONICET,
Argentina.} Julio F. Navarro,$^{1,3}$\thanks{Fellow of CIAR and
of the J. S. Guggenheim Memorial Foundation.} 
and Matthias Steinmetz$^{2}$
\\
$^{1}$Department of Physics and Astronomy, University of Victoria, Victoria, BC V8P 5C2,
Canada\\
$^{2}$Astrophysikalisches Institut Potsdam, An der Sternwarte 16, Potsdam 14482, Germany\\
%
%
$^{3}$Max-Planck Institut f\"ur Astrophysik, Karl-Schwarzschild Strasse 1,
Garching bei M\"unchen, D-85741, Germany\\
}

\date{}
\pubyear{2005}
\maketitle
\label{firstpage}

\begin{abstract}
We use numerical simulations to investigate the origin and structure of the
luminous halos that surround isolated galaxies. These stellar structures
extend out to several hundred kpc away from a galaxy, and consist of stars
shed by merging subunits during the many accretion events that characterize
the hierarchical assembly of galaxies. Such origin suggests that outer
luminous halos are ubiquitous and that they should appear as an excess of
light over extrapolations of the galaxy's inner profile beyond its traditional
luminous radius. The mass profile of the accreted stellar component is well
approximated by a model where the logarithmic slope steepens monotonically
with radius; from $\rho \propto r^{-3}$ at the luminous edge of the galaxy to
$r^{-4}$ or steeper near the virial radius of the system. Such spatial
distribution is consistent with that of Galactic and M31 globular clusters,
suggesting that many of the globulars were brought in by accretion events, in
a manner akin to the classic Searle-Zinn scenario. Luminous halos are similar
in shape to their dark matter counterparts, which are only mildly triaxial and
much rounder than dark halos formed in simulations that do not include a
dissipative luminous component. The outer stellar spheroid is supported by a
velocity dispersion tensor with a substantial and radially increasing radial
anisotropy; from $\sigma_r^2/\sigma_t^2 \sim 2$ at the edge of the central
galaxy to $\sim 5$ at the virial radius. These properties distinguish the
stellar halo from the dark matter component, which is more isotropic in
velocity space, as well as from some tracers of the outer spheroid such as
satellite galaxies. Most stars in the outer halo formed in progenitors that
have since merged with the central galaxy; very few stars in the halo are
contributed by satellites that survive as self-bound entities at the
present. These features are in reasonable agreement with recent observations
of the outer halo of the Milky Way, of M31, and of other isolated spirals, and
suggest that all of these systems underwent an early period of active merging,
as envisioned in hierarchical models of galaxy formation.
\end{abstract}

\section{Introduction}
\label{sec:intro}
Galaxies have no edge. With rare exceptions, the stellar spatial distribution
in normal galaxies shows little sign of a sharp outer cutoff, and is
reasonably well approximated by density laws that extend smoothly to
arbitrarily large radius. Extrapolations of the inner luminosity profile,
however, suggest that little light comes from regions of surface brightness
much fainter than those traditionally used to define the luminous radii of
galaxies ($\sim 25$ mag/asec$^2$). Perhaps because of this reason, together
with the obvious observational difficulties inherent to studying regions of
low-surface brightness, the outer luminous halos of external galaxies have in
the past been regarded as a topic of little more than academic interest.

This impression, however, is rapidly changing, as new datasets start to unveil
some unexpected properties of the stellar component that populates the outer
confines of galaxies. These developments have been made possible by the
development of panoramic digital cameras able to map the light distribution of
external galaxies down to unprecedented surface brightness levels,
complemented by efficient observational techniques designed to identify and
measure radial velocities of outer halo tracers in external galaxies, such as
planetary nebulae (PNe, Romanowsky et al. 2003). Finally, the advent of
efficient spectrographs in 10m-class telescopes have enabled the measurement
of radial velocities of large samples of giant stars throughout the Local
Group (Ibata et al 2004), and dedicated spectroscopic campaigns targeted on
giant stars have dramatically increased the sample of tracers in the outer
halo of our own Milky Way (Morrison et al 2003, Battaglia et al 2005, Clewley
et al 2005).

Likewise, the use of stacking techniques in wide-field surveys such as the
Sloan Digital Sky Survey (SDSS) have led to clear detections of a distinct low
surface brightness component around both isolated spiral galaxies (Zibetti,
White \& Brinkmann 2004) as well as central cluster galaxies (Zibetti et al
2005). The surface brightness profile of these outer luminous halos deviate
significantly from straightforward extrapolations of the laws that describe
the main body of the galaxy, and suggest that estimates of the amount of light
in the intergalactic or intracluster medium might need to be revised
upward. In the case of clusters, for example, Zibetti et al (2005) conclude
that of order 11\% of the light in a typical Abell cluster might be in the
form of an intracluster luminous component with structural properties distinct
from those of the galaxy cluster population and of the central cluster galaxy.

These studies have brought about several unexpected results that challenge
some of the accepted premises of the currently accepted galaxy formation
paradigm. For example, the kinematics of PNe in the vicinity of several normal
ellipticals has been found to be consistent with simple models where such
galaxies have no surrounding dark matter halo (Romanowsky et al. 2003, but see
Dekel et al 2005). This result, if confirmed by further studies, would be
very difficult to accommodate in the $\Lambda$CDM scenario, where {\it all}
galaxies are envisioned to form as a result of the dissipative collapse of
baryons within massive dark matter halos.

In the case of the Local Group, wide field photometric surveys (and their
follow-up spectroscopic campaigns) have modified radically the traditional
view of the outer stellar spheroid of the Milky Way and M31.  Giant structures
interpreted as tidal streams of disrupted or disrupting dwarfs have been
identified in both galaxies (Ibata et al 1994, Helmi et al 1999, Majewski et
al 2003, Ibata et al 2004), and have solidified the notion that accretion
events may play a substantial role in the shaping of individual
galaxies. These streams are expected to be short-lived as readily identifiable
structures on the sky (Johnston et al 1995, Helmi \& White 1999), and,
although impressive, they are thought to contribute a relatively small
fraction of stars in the halo of these systems. In broad terms, the main
characteristics of the outer stellar halo are probably adequately captured by
a simple model of a reasonably well-mixed spheroid of stars.

Extensive spectroscopic surveys of giant stars in the outskirts of Local Group
galaxies have unraveled some intriguing dynamical properties for the outer
halo component. For example, just like PNe around ellipticals, the velocity
dispersion of these halo tracers drops significantly in the outer regions of
the Milky Way (Battaglia et al 2005, Clewley et al 2005). This is a somewhat
unexpected result if these galaxies are embedded in extended dark matter
halos, and has prompted renewed interest in the extent and spatial
distribution of dark matter in the outskirts of galaxies.

From a theoretical point of view, the interest elicited by outer luminous
halos stems from the fact that, in the currently accepted paradigm, stars are
envisioned to form only in the collapsed, high-density regions near the center
of dark halos---which we identify with the main body of individual
galaxies. This prejudice is supported by observations that indicate the need
for a threshold gas density below which stars do not form (Kennicutt 1989,
Martin \& Kennicutt 2001, Schaye 2004), and imply that outer halo stars did
{\it not} form {\it in situ} but have been shed from protogalaxies during the
merger events that characterize the assembly of galaxies in a hierarchically
clustering universe.

In simple words, stars found as far away as, say, 100 kpc from a galaxy's
center originate in satellites whose orbital apocenter was about that large
before they spiraled in to merge with the main galaxy. We therefore expect a
clear connection between the orbital properties of stars in the outer halo and
those of the progenitors that merged to form the present-day
galaxy. Unraveling the structure of the outer luminous halo of galaxies may
thus be seen as an important step toward unraveling the merging history of
individual galaxies.

In this paper, we analyze the origin and structure of the luminous halos of
galaxies simulated in the $\Lambda$CDM scenario.  We describe briefly in
Section \ref{sec:numexp} the numerical simulations.  The main results of our
analysis are presented in Section \ref{sec:results} and discussed in
\S~\ref{sec:disc}. We conclude with a brief summary in Section \ref{sec:conc}.

\begin{figure*}
\begin{center}
\includegraphics[width=\linewidth,clip]{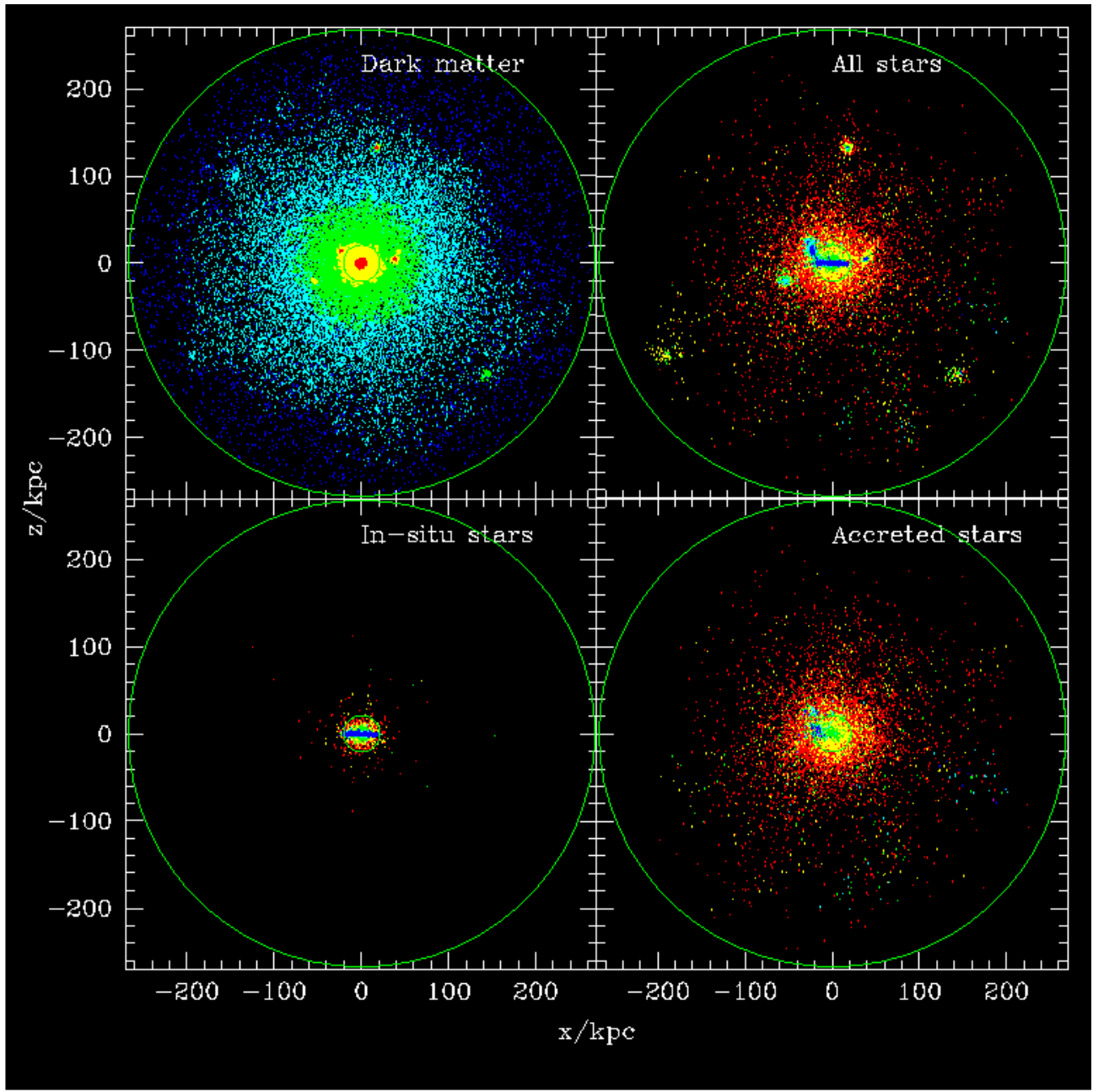}
\end{center}
\caption{Spatial distribution of the dark matter and stellar component of one
simulated galaxy at z=0. Top panels show the dark matter (left) and the stars
(right) within the virial radius of the system. Dark matter particles are
colored by their local density, whereas stars are colored by age (blue: 0.0
$<$ age/Gyr $<$ 2.5, cyan: 2.5 $<$ age/Gyr $<$ 5.0, green: 5.0 $<$ age/Gyr $<$
7.5, yellow: 7.5 $<$ age/Gyr $<$ 10.0, red 10.0 $<$ age/Gyr $<$ 15.0). The
outer green circle shows the virial radius of the system, the inner circle
shows the radius adopted as the ``luminous radius'' of the galaxy. Bottom
panels split the stellar component into two different groups: stars formed in
the main progenitor of the galaxy (i.e., ``in situ'' stars, left panel) and
those contributed by accretion events (``accreted'' stars, right panel). The
latter group excludes stars in satellites that remain self-bound in the halo
of the galaxy.
\label{figs:xypanel}}
\end{figure*}

\section{The Numerical Simulations}
\label{sec:numexp}

We analyze a suite of eight numerical simulations of the formation of galaxies
in the $\Lambda$CDM scenario. These simulations have been analyzed in earlier
papers, which may be consulted for details on the code used as well as on the
numerical setup (Steinmetz \& Navarro 2002, Abadi et al 2003a,b, Meza et al
2003, 2005). In brief, each simulation follows the evolution in a $\Lambda$CDM
universe of a small region surrounding a target galaxy, excised from a large
periodic box and re-simulated at higher resolution preserving the tidal fields
from the whole box. The simulation includes the gravitational effects of dark
matter, gas and stars, and follows the hydrodynamical evolution of the gaseous
component using the Smooth Particle Hydrodynamics (SPH) technique (Steinmetz
1996). We adopt the following cosmological parameters for the $\Lambda$CDM
scenario: $H_0=65$ km/s/Mpc, $\sigma_8=0.9$, $\Omega_{\Lambda}=0.7$,
$\Omega_{\rm CDM}=0.255$, $\Omega_{\rm bar}=0.045$, with no tilt in the
primordial power spectrum. All simulations start at redshift $z_{\rm
init}=50$, have force resolution of order $1$ kpc, and mass resolution so that
each galaxy is represented, at $z=0$, with $\sim 125,000$ star particles.

Dense, cold gas in collapsed regions is allowed to turn into stars at rates
consistent with the empirical Schmidt-like law of Kennicutt (1998). The
energetic feedback of evolving stars is included mainly as a heating term on
the surrounding gas, but the short cooling times in these regions reduce
significantly the effectiveness of feedback in curtailing star formation. The
transformation of gas into stars thus tracks closely the rate at which gas
cools and condenses at the center of dark matter halos. As discussed in the
references listed in the above paragraph, this results in an early onset of
star-forming activity in the many progenitors of the present-day galaxy
present at high redshift. This leads to the formation of a prominent
spheroidal component as these progenitors coalesce and merge to form the final
galaxy. Gas accreted after the merging activity abates leads to the formation
of a centrifugally-supported gaseous and stellar disk component clearly
present in most of our simulations, although they make up typically only about
$\sim 30\%$ of the final stellar mass of each galaxy.

It is likely that improvements to our feedback algorithms may lead to
revisions in the efficiency and timing of star formation in these galaxies. As
we discuss below, outer halo stars originate mainly in the merger of the early
progenitors, so reducing the efficiency of early star formation should have
some impact on the actual prominence of the outer stellar halo component. We
expect, however, that reasonable modifications of the star formation algorithm
will affect mainly the number, ages, and chemical composition of outer halo
stars, rather than its dynamical and structural properties, which we expect to
depend mainly on the orbital properties of the merging progenitors and on the
structure of the dark matter halo host. These properties are less sensitive to
the complex astrophysics of star formation and feedback, and we therefore
focus our analysis on the structure and dynamics of the outer stellar halo in
our eight simulations.  These target a small range in mass chosen so that at
$z=0$ the galaxies have luminosities and circular speeds comparable to the
Milky Way.  We summarize the main structural parameters of the dark matter and
stellar components of the simulated galaxies in Table~\ref{tab:table1}.

\begin{figure}
\begin{center}
\includegraphics[width=\linewidth,clip]{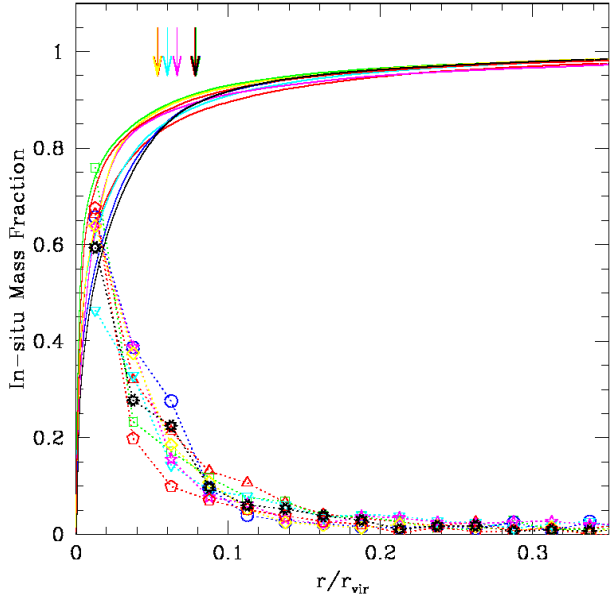}
\end{center}
\caption{The radial dependence of the fraction of stars formed in the most
  massive progenitor of the final galaxy (i.e., stars formed {\it in situ}) is
  indicated by dotted lines connecting symbols. The thick solid lines indicate
  the cumulative fraction of stars within the virial radius. In-situ stars
  dominate throughout the main body of the galaxy; within the ``luminous
  radius'' of the galaxy, defined to be $r_{\rm lum}=20$ kpc (shown with
  arrows), they contribute $\sim 60\%$ of the stellar mass. Outside $r_{\rm
  lum}$ the stellar component consists almost exclusively of accreted stars;
  indeed, fewer than $\sim 5\%$ of stars in the outer ($r>r_{\rm lum}$) halo
  formed in situ.
\label{figs:rfinsitu}}
\end{figure}

\begin{figure}
\begin{center}
\includegraphics[width=\linewidth,clip]{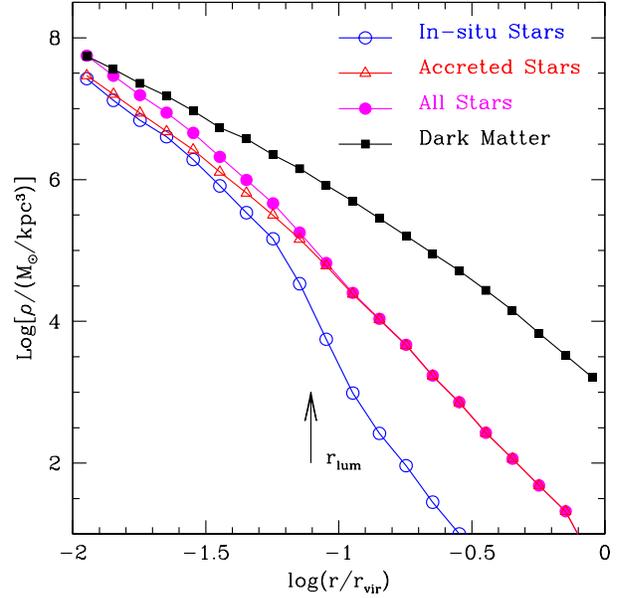}
\end{center}
\caption{Density profile of stars (circles) and dark matter (squares) in one
  of our simulations (KIA3), shown in a logarithmic scale. Open circles
  correspond to the stars formed in situ, whereas filled circles correspond to
  accreted stars. Note that the {\it shape} of the density profile of the
  accreted stellar component is similar to that of the dark matter, and can be
  adequately fit with a radial law where the logarithmic slope is a power law of
  radius. See text for details.
\label{figs:rhoprof}}
\end{figure} 

\section{Results}
\label{sec:results}

Figure~\ref{figs:xypanel} shows, at $z=0$, one of our simulated galaxies
(KIA3, see Table~\ref{tab:table1}) projected onto a box of $540$ kpc on a
side. The top panels show the dark matter particles (left) and stars (right)
within the virial radius ($r_{\rm vir}\approx 270$ kpc, shown by the outer
green circle), defined to encompass a region of mean density $100$ times the
critical density for closure. Dark matter particles are coloured by their
local density, while stars are coloured by their age, as described in the
label.

The bottom panels in Figure~\ref{figs:xypanel} separate the stars in two
components: ``in situ'' stars that formed in the most massive progenitor
(left) and ``accreted stars'' that formed in progenitors that merged with the
main galaxy (right). Stars labelled as ``accreted'' exclude those associated
with self-bound satellite systems that survive until the present. These can be
seen clearly in the top-right panel of Figure~\ref{figs:xypanel}, but are
largely absent in the bottom-right panel, except for a sprinkling of stars
associated with a young stream recently torn from a disrupting satellite,
visible at $x\sim -20, y\sim 10$ kpc. Roughly $\sim 48\%$ of stars (by mass)
formed ``in situ'' in this galaxy, compared with $\sim 44\%$ which make up the
accreted component. Satellites contribute a rather small fraction ($\sim 8\%$)
of all stars within $r_{\rm vir}$. These numbers are typical of our
simulations, as may be seen from the numbers listed in Table~\ref{tab:table1}.

\begin{figure}
\begin{center}
\includegraphics[width=\linewidth,clip]{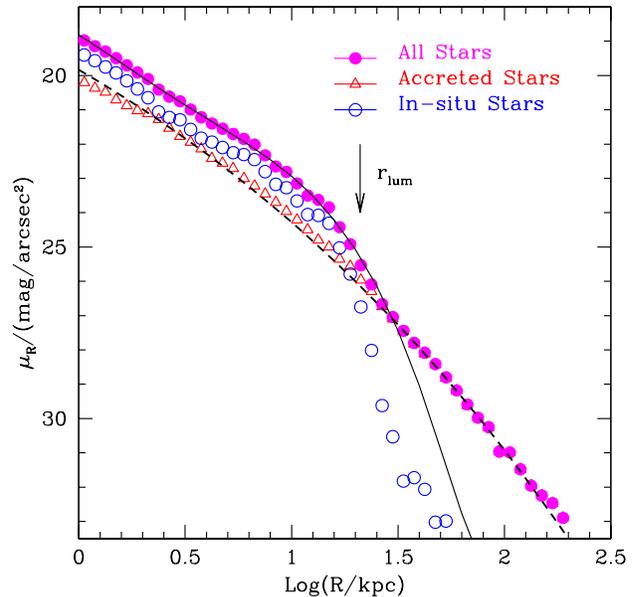}
\end{center}
\caption{$R$-band surface brightness profile of stars, split between the
  ``in-situ'' and ``accreted'' components (run KIA3, see
  Table~\ref{tab:table1}). The accreted stars dominate outside the luminous
  radius and appear as an excess of light over a bulge+disk fit to the inner
  surface brightness profile, which is shown as a solid line. The dashed line
  shows a Sersic-law fit to the outer profile of accreted stars. Note that a
  single Sersic-law (with parameters given in Table~\ref{tab:table2})
  reproduces very well the radial distribution of all accreted stars.
\label{figs:sbprof}}
\end{figure} 

Figure~\ref{figs:xypanel} illustrates a number of properties of the stellar
component common to all of our simulations. In particular, it is important to
note that (i) stars spread as far out as the virial radius of the system
(outer circle in the panels of Figure~\ref{figs:xypanel}), although they are
more highly concentrated than the dark matter halo; (ii) {\it in situ} stars
(bottom left panel), are responsible for most of the young stars in the main
body of the galaxy (note the prominence of the young disk in the bottom-left
panel), and are practically absent from the outer halo; whereas (iii) accreted
stars (bottom right panel) make up preferentially the spheroidal component and
dominate the stellar budget in the outer regions of the galaxy.

The relative importance of the ``accreted'' vs ``in situ'' components at
various radii is shown in Figure~\ref{figs:rfinsitu}. Beyond a radius $r_{\rm
lum}=20$ kpc (marked by arrows in Figure~\ref{figs:rfinsitu}) accreted
stars dominate in all our simulations. We shall define $r_{\rm lum}$ as the
``luminous radius'' of the galaxy and refer to stars beyond $r_{\rm
lum}$ as the ``outer luminous halo'' or ``outer galaxy'', for short. Our
analysis focuses on the structure and dynamics of the outer halo and,
consequently, mainly on the properties of the accreted component. 

It is also clear from this figure that the outer luminous halo contains a
relatively small fraction of the stellar mass of the galaxy; fewer than 15\%
of all stars are found between $r_{\rm lum}$ and the virial radius.  We note
as well that accreted stars make up a non-negligible fraction of stars in the
inner galaxy (i.e., inside $r_{\rm lum}$); we shall therefore distinguish
between ``accreted'' and ``outer halo'' stars in what follows.

Figure~\ref{figs:rhoprof} shows the density profile of the stellar and dark
matter components of the simulation shown in Figure~\ref{figs:xypanel}. The
stellar contribution is split between in-situ and accreted stars. Note that
stars are significantly more concentrated than the dark matter; indeed, they
contribute a significant fraction of the {\it total} mass within the luminous
radius ($44\%$ on average for all runs in our series), but they make up a
negligible amount of the total mass in the outer regions. Note as well the
sharp truncation of in-situ stars at $r>r_{\rm lum}$; essentially no stars in
the outer halo were formed in the main progenitor. This is a natural
consequence of the fact that in-situ star formation proceeds efficiently only
in regions of high-density. The few in-situ stars found beyond $r_{\rm lum}$
at $z=0$ have been torn from the main progenitor during mergers, although
Figure~\ref{figs:rfinsitu} (and the data in Table~\ref{tab:table1}) shows that
this is rather inefficient, and that the ``accreted'' component dominates the
stellar component of the outer halo.

\begin{figure*}
\begin{center}
\includegraphics[width=0.475\linewidth,clip]{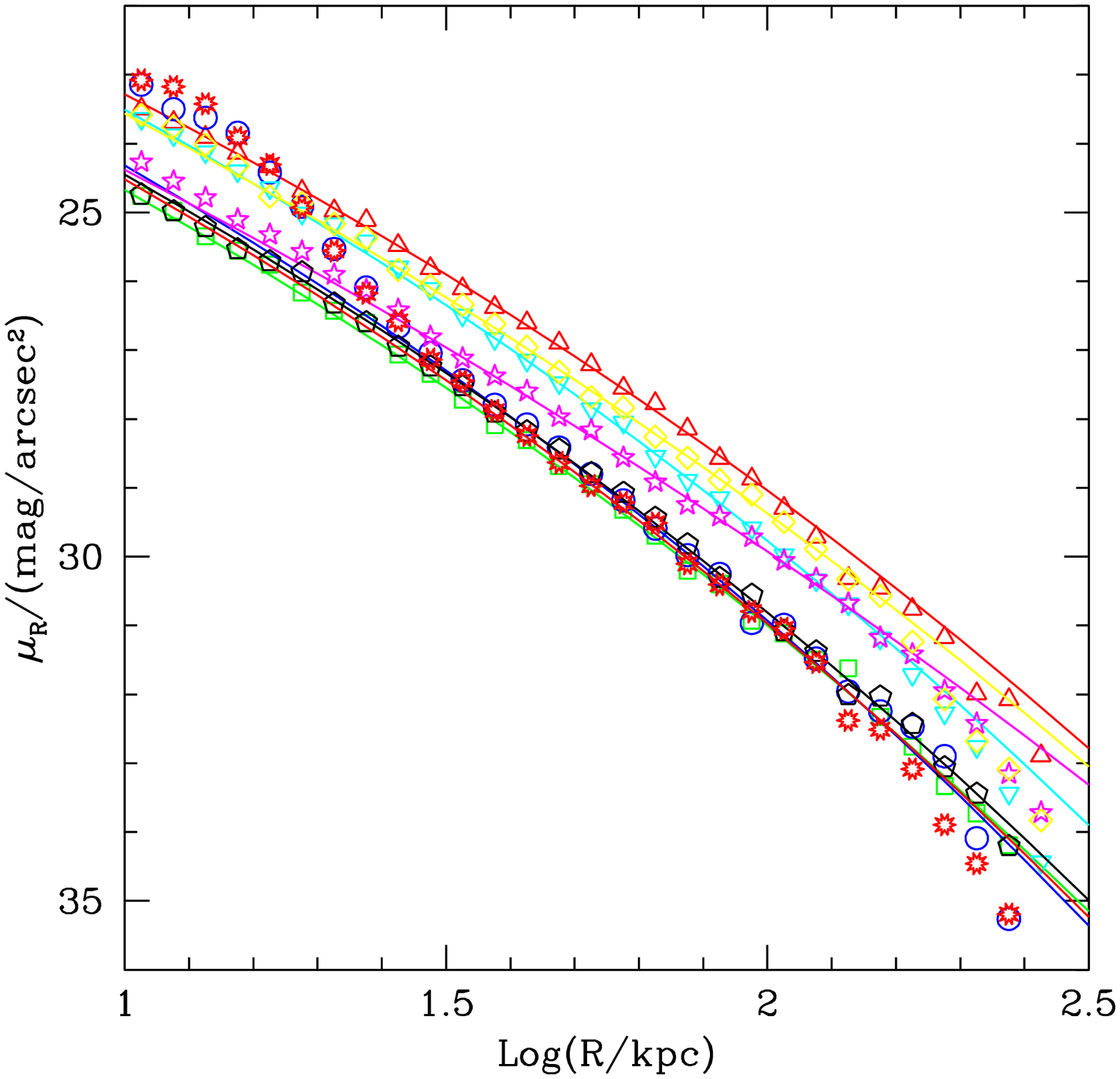}
\hspace{0.4cm}
\includegraphics[width=0.475\linewidth,clip]{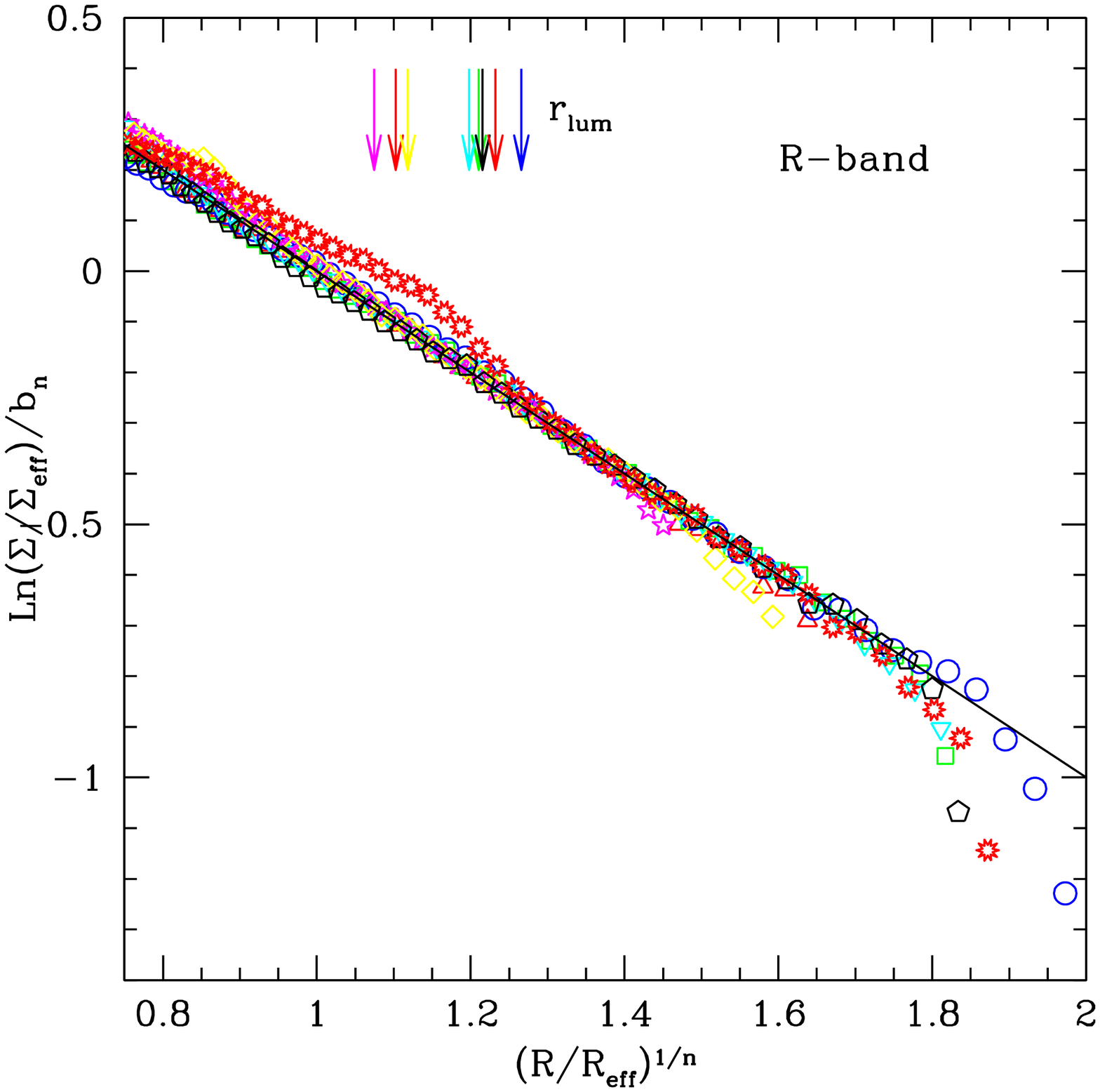}
\end{center}
\caption{({\it a-left}) R-band surface brightness profile of the outer galaxy
  ($10<R/$kpc$<300$) in all of our simulations. Solid lines correspond to the
  best Sersic-law fits to the outer regions of the profile ($r>r_{\rm lum}$)
  whose parameters are listed in Table~\ref{tab:table2}. ({\it b-right}) Same
  as in the left, but profiles correspond {\it only to accreted stars} and
  have been scaled as stated in the axis labels so that Sersic-law profiles
  would follow the straight solid line of slope $-1$. Interestingly, a simple
  Sersic-law reproduces quite well the radial distribution of accreted stars
  in all simulations. There is one exception (which shows as a bump in the
  right-hand panel), associated with the transient effect of an ongoing
  satellite disruption event.
\label{figs:sersic}}
\end{figure*} 

The radius beyond which accreted stars begin to dominate is signaled by an
abrupt change in the surface brightness profile of the galaxy, as shown in
Figure~\ref{figs:sbprof}. Inside the luminous radius, where
$\mu_R$ is brighter than about $25.5$ mag/asec$^2$, the surface brightness
profile of this galaxy (projected in this case face-on and studied in detail
by Abadi et al 2003a,b) is well approximated by a de Vaucouleurs bulge plus an
exponential disk (shown as a solid line). This bulge+disk model is, however,
unable to fit the structure of the outer luminous halo, which deviates from
the inner profile and appears as a outer luminous ``excess'' beyond $\sim 30$
kpc, where the $R$-band surface brightness drops to values fainter than about
$26.5$ mag/asec$^2$.

The outer halo surface brightness profile steepens gradually outwards and is
well approximated by a Sersic-law, 
\begin{equation}
\Sigma(R) = \Sigma_{\rm eff} \, e^ { -b_n [(R/R_{\rm eff})^{1/n}-1]}
\end{equation}
where $R_{\rm eff}$ and $\Sigma_{\rm eff}$ are, respectively, the radius
containing half the light and the surface brightness level at that radius.
The surface brightness profile of the outer regions of our simulated galaxies
is shown in Figure~\ref{figs:sersic}a, together with Sersic-law fits to the
outer portion of the profiles.

The Sersic fit parameters are listed in Table~\ref{tab:table2}. On average we
find $\langle n \rangle =6.3$; $\langle R_{\rm eff} \rangle =7.7$ kpc. The
outer luminous halo's surface brightness profile steepens from $\Sigma(R)
\propto R^{-2.3}$ at about the luminous radius to $R^{-2.9}$ at $r\sim 100$
kpc and to $R^{-3.5}$ or steeper around the virial radius. The gradually
steepening profile of the outer halo characterized by the Sersic law is
reminiscent of the distribution of dark matter, whose density profile also
steepens monotonically outward (Navarro et al 2004). The dark matter profile
is, however, much less centrally concentrated and its slope is shallower than
the stellar halo at all radii. As discussed by Merritt et al (2005), the dark
matter density profile may also be approximated by a Sersic law, but with a
characteristic value of $2 < n < 4$ and much larger effective radii.

Figure~\ref{figs:sersic}b shows the surface brightness profiles {\it of the
accreted stellar component only} for all eight simulations in our series. The
profiles have been scaled so that they would all line up along a line of slope
-1 if they followed accurately a Sersic law. All profiles are indeed well
approximated by eq.~1, with the possible exception of one case where a
``bump'' of stars associated with an ongoing satellite disruption event is
seen just outside $R_{\rm eff}$ (open starred symbols in
Figure~\ref{figs:sersic}b). This raises the interesting prospect of estimating
the {\it total} fraction of stars accreted throughout the history of a galaxy
simply by fitting its outer profile. As the last two columns of
Table~\ref{tab:table2} demonstrate, the total light contributed by accreted
stars is well approximated (to within a factor of $\sim 2$) by the total light
of a Sersic-law fit to the outer luminous halo.

The similarity between profiles suggest that the violent relaxation associated
with mergers endows the accreted stellar component with a simple, roughly
self-similar structure that is well approximated by a Sersic law. This process
has been studied extensively in the literature, and previous studies have
consistently found that the structure of N-body merger remnants is quite
reminiscent to that of bright ellipticals (see, e.g., Barnes \& Hernquist 1992
and references therein), which are indeed well approximated by a Sersic law
with $n\gsim 4$ (see, e.g., Graham \& Guzman 2003 and references therein).

\begin{figure}
\begin{center}
\includegraphics[width=\linewidth,clip]{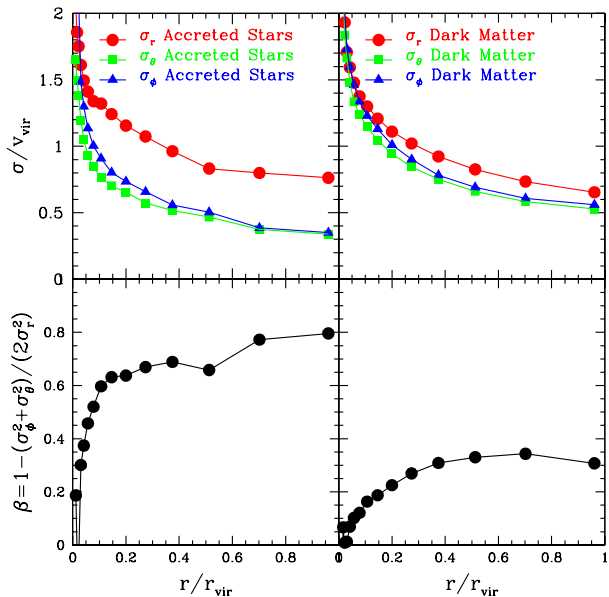}
\end{center}
\caption{Average velocity dispersion profile of stars and dark matter in our
  simulations. Averages are computed by scaling all positions and velocities
  to the virial radius and virial circular velocity of each system,
  respectively. The rotation axis of the inner galaxy is chosen as the polar
  axis of the scaling procedure. Particles are then grouped in spherical bins
  and the spherical components of the velocity dispersion are computed.  These
  profiles are finally averaged over all systems to produce the radial and
  tangential velocity dispersion profiles plotted here.  Right panels
  correspond to the dark matter component and left panels to the accreted
  stars. Recall that, on average, the luminous radius is $\sim 0.06 \, r_{\rm
  vir}$ in these units. Note that the dark matter velocity dispersion exhibits
  a mild radial bias; the radial bias is much more pronounced, and
  monotonically increasing, in the case of the accreted stellar component.
\label{figs:sigmar}}
\end{figure}

One may think of the accreted stellar component as formed by the overlap of
the many ``tidal tails'' stripped from the merging progenitors; in this
interpretation the outer luminous halo would just be the superposition of the
outer tails stripped from each progenitor during the merger process.  This
characterization of the accreted stellar component helps to explain the
remarkable kinematics of the outer halo shown in
Figure~\ref{figs:sigmar}. This figure shows the velocity dispersion profile of
stars and dark matter, averaged over all simulations after scaling the
positions and velocities of all particles to the virial radius and virial
velocity of each system. The right panels in Figure~\ref{figs:sigmar} show
that the velocity dispersion of the dark matter declines gradually from the
center outwards, and is characterized by a mild radial anisotropy, going from
almost isotropic near the center to $\beta=(1-\sigma_t^2/\sigma_r^2) \sim 0.3$
in the outer regions. (We use $\sigma_t$ to denote the tangential velocity
dispersion; $\sigma_t^2=(\sigma_{\theta}^2+\sigma_{\phi}^2)/2$.)

The outer luminous halo, on the other hand, shows a much more pronounced
radial bias: from $\beta\sim 0.4$ just outside the luminous radius of the
galaxy ($r_{\rm lum}$ is on average of order $0.06 \, r_{\rm vir}$) to about
$\beta\sim 0.8$ near the virial radius. This substantial--and monotonically
increasing--radial anisotropy may also be understood as a direct consequence
of the tidal disruption process of formation described above. Outer halo stars
are typically stripped from merging progenitors during pericentric
passages. However, because stars form in high-density regions the stripping of
stars operates efficiently only when pericenters are small (and tidal effects
are greatest). This occurs typically after dynamical friction has eroded the
orbit sufficiently to bring the pericenter of the accreting satellite close to
the central galaxy. As a result, outer halo stars are ``launched'' into highly
energetic orbits with, characteristically, the (small) pericentric radii that
accompany the most disruptive tides. Stars able to reach farther are thus
typically on more eccentric orbits, leading to the increasing radial
anisotropy in the orbits of the outer halo stars seen in
Figure~\ref{figs:sigmar}.

\begin{figure}
\begin{center}
\includegraphics[width=\linewidth,clip]{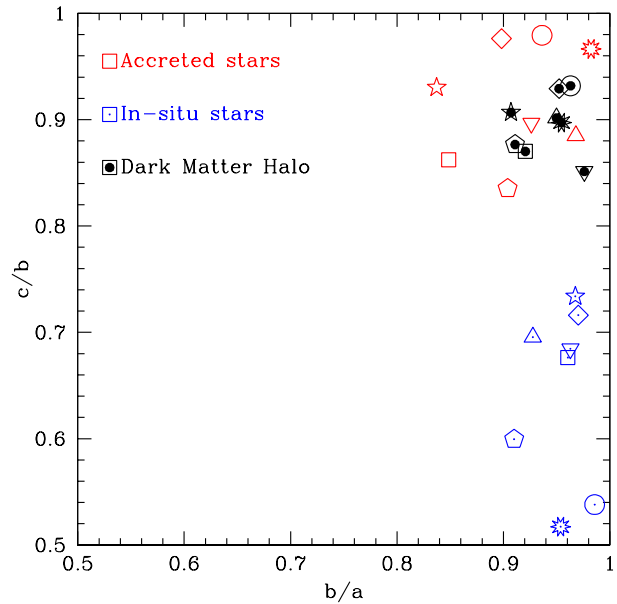}
\end{center}
\caption{Inertia axis ratios of the dark matter halo, as well as of the
  accreted and in-situ stellar components. Axis ratios are computed by
  diagonalizing the inertia tensor $I_{ij}=\sum m_i x_i x_j $ for all
  particles of each component within the virial radius of the system. Note
  that the shape of the dark matter and accreted components are similar and
  only mildly triaxial, but that the in-situ stars are predominantly
  oblate. This reflects the fact that in-situ stars have a higher proportion
  of newly formed stars, which tend to arrange themselves in a disk-like
  structure (see Figure~\ref{figs:xypanel}).
\label{figs:cbba}}
\end{figure}

The formation process of the accreted stellar component is thus qualitatively
similar to that of the dark matter halo, where mergers also play a substantial
role. It is thus perhaps not surprising that they settle onto similar, mildly
triaxial structures, as shown in Figure~\ref{figs:cbba}. For the accreted
stars, the average intermediate-to-major axis ratio is $\langle b/a \rangle
=0.91$, whereas the average minor-to-intermediate axis ratio is $\langle c/b
\rangle =0.92$ ($\langle c/a \rangle =0.84$). This is not too different from
the dark matter halos, which have on average $ \langle b/a \rangle =0.94$ and
$ \langle c/b \rangle = 0.90$ ($ \langle c/a \rangle=0.84$).  We note that the
dark matter halos in these simulations are much rounder than typically found
in N-body simulations of CDM halo formation: $ \langle b/a \rangle \sim 0.75$
and $ \langle c/a \rangle \sim 0.6$ (see, for example, Bailin \& Steinmetz
2004). This is the result of the response of the halo to the dissipative
assembly of the baryonic component of the galaxy at its center, which steepens
the potential well and reduces the triaxiality of the halo; see, e.g.,
Kazantzidis et al (2004), Bailin et al (2005), and references therein.

The in-situ stellar component, on the other hand, often sports a well defined
rotationally-supported disk (see, e.g., lower left panel of
Figure~\ref{figs:xypanel}) and its shape is thus well described by an oblate
structure with $\langle b/a \rangle=0.95$ and $\langle c/a
\rangle=0.62$. Although rotation plays little role supporting the accreted
component (the mean rotation velocity, ${\bar v}_{\phi}=j_z/R$, is typically
less than one tenth of the circular velocity at each radius), it is
interesting to note that the direction of its angular momentum is well aligned
with that of the inner galaxy.

\begin{figure}
\begin{center}
\includegraphics[width=\linewidth,clip]{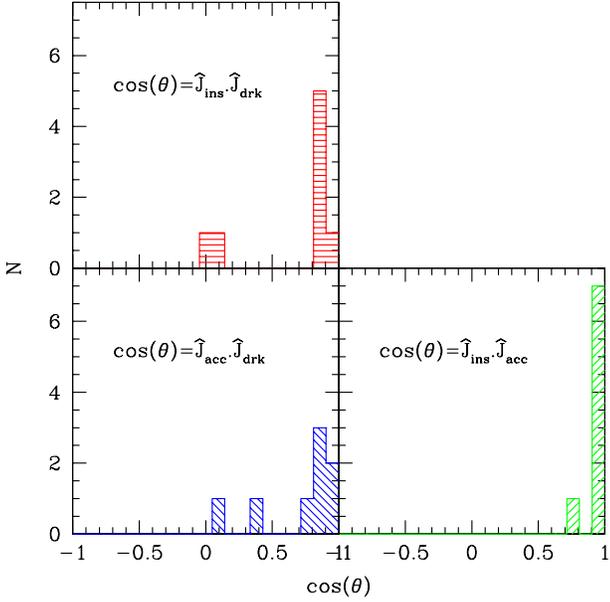}
\end{center}
\caption{ Distribution of the cosine of the angle between the angular momentum
vector of the in-situ stellar component, ${\vec J}_{\rm ins}$, the accreted
stars, ${\vec J}_{\rm acc}$, and the dark matter, ${\vec J}_{\rm drk}$. Each
of the three different permutations are shown by different shaded histograms,
as indicated in the labels of each panel.
\label{figs:jjdist}}
\end{figure}

This is shown in Figure~\ref{figs:jjdist}, which show the distribution of the
cosine of the angles between the angular momentum of the dark matter halo
(${\vec J}_{\rm drk}$), the in-situ stars (${\vec J}_{\rm ins}$), and the
accreted stellar halo (${\vec J}_{\rm acc}$). Although the good agreement
between ${\vec J}_{\rm acc}$ and ${\vec J}_{\rm ins}$ could have perhaps been
anticipated (after all, they are both part of the same stellar system) it is
still interesting to note the strong correlation between the rotational
properties of the inner galaxy and the dark matter halo. This implies that,
although mergers lead to substantial transfer of angular momentum from the
stars to the dark halo (Navarro \& Steinmetz 1997, Abadi et al 2003a), this
does not alter radically its orientation, so that the spin of the inner galaxy
generally ends up aligned with the rotation axis of the surrounding dark
matter halo.

It is also of interest to characterize the orientation of the rotation axis
relative to the principal axes of the system. Figure~\ref{figs:jcdist} shows
the alignment between the galaxy's angular momentum and the minor axis of the
dark matter and luminous halos.  No obvious alignment is seen in these panels,
suggesting that the rotational properties of the galaxy are, at best, weakly
correlated with the shape of the surrounding dark matter (or luminous)
halo. We shall return to possible interpretations of this result in the
following section.

We end the characterization of the outer luminous halos of simulated galaxies
by comparing the age distribution of stars in the inner and outer galaxy, as
well as that of satellites enclosed within the virial radius
(Figure~\ref{figs:agedist}). The ages of stars in the outer galaxy ($r>r_{\rm
lum}$) differ significantly from those in higher density regions (like the
inner galaxy or the surviving satellites), where star formation may
proceed. The outer halo is populated mainly by older stars, reflecting the
fact that the mergers responsible for its formation are more common at earlier
times. Interestingly, the distribution of ages of stars in the outer halo is
also fairly distinct from that of stars in satellites orbiting within the
virial radius. This shows that relatively few stars in the outer halo
originate in the ``harassment'' of satellites that have survived as self-bound
entities until the present (Moore et al 1996). Most stars in the outer halo
come from merger events whose progenitors have long been fully disrupted,
suggesting that the properties of the satellite population may be quite
distinct from that of the smooth outer halo. We explore the consequences of
these results for the interpretation of observational data in the following
section.

\section{Discussion}
\label{sec:disc}

The results presented in the previous section have a number of implications
regarding the interpretation of observations of stars in regions far removed
from the normal boundaries of typical galaxies. Our simulations suggest that
these ``intergalactic stars'' (or, more properly, ``extra-galactic stars'')
are a relict of the merging history of each individual galaxy. As such, all
galaxies with a past of active merging activity (the majority in a
hierarchically clustering scenario such as $\Lambda$CDM) should have an outer
stellar halo component with properties similar to those described in the
previous section.

The outer luminous halo should appear as an excess of light over the
extrapolated profile of the inner galaxy. Its structure is characterized by a
gradually steepening density profile and by a clearly defined radial
anisotropy in the velocity distribution.  Outer halos are rather difficult to
detect, since they are--by definition--confined to regions of very low surface
brightness difficult to probe observationally. However, the presence of outer
luminous components detached from the properties of the inner galaxy have been
recently reported by a number of authors. Following the pioneering work of
Sackett et al (1994), Morrison et al (1994), and Zheng et al (1999) among
others, Zibetti, White \& Brinkmann (2004), have recently reported the
statistical detection of a $\rho \propto r^{-3}$ halo of stars in the
outskirts of a sample of edge-on disk-dominated galaxies selected from the
Sloan Digital Sky Survey (SDSS).

Encouragingly, this outer halo component is detected as excess light over an
extrapolation of the inner disk (see Figure~\ref{figs:sbprof}); the halo
dominates the minor axis light profile beyond about $\sim 6$ exponential
scalelengths, in regions where the $r$-band surface brightness drops to values
fainter than $\mu_r\sim 27.5$. A qualitatively similar result has been derived
for M31 through detailed stellar counts by Guhathakurta et al (2005) and by
Irwin et al (2005), who find that the surface brightness profile along the
minor axis ``flattens'' relative to an extrapolation of the de Vaucouleurs fit
to the inner spheroid at radii beyond $\sim 20$ kpc.

Our simulations provide a compelling interpretation where outer stellar halos
are made up predominantly of stars accreted during previous merger events and,
in particular, by the subset which were propelled into highly energetic (and
highly eccentric) orbits during the disruption process that accompanies the
mergers. Mergers and phase mixing lead then to the formation of a tenuous,
distinct stellar component that fills the halo of the galaxy out to the virial
radius.

Although we expect them to be ubiquitous, the prominence of these outer
luminous halos is difficult to assess from a theoretical point of view, as it
depends critically on the number and timing of merger events, as well as on
the fraction of stars present at the time the mergers occur, all of which
remain highly uncertain. A number of general inferences are, however, still
possible. Late major mergers should propel a proportionally larger fraction of
stars into the outer halo, and therefore the relative prominence of this
component should be higher in spheroid-dominated galaxies, which are widely
regarded as merger remnants. This process should be especially important in
systems that have undergone repeated mergers, leading to the expectation that
the relative importance of the ``intracluster'' stellar component should be
highest in clusters with prominent cD galaxies.

In disk-dominated galaxies, where in-situ star formation is typically still
ongoing and mergers are less important, the outer halo is likely to dominate
the profile only at large radii, where star formation thresholds prevent the
efficient formation of in-situ stars (Kennicutt 1989, Ferguson et al 1998,
Schaye 2004).  The accreted stellar component---to which no such threshold
applies---appears then as a distinct outer component, ``flattening'' the
surface brightness profile in the outer regions, as reported by Zibetti et al
(2004), Guhathakurta et al (2005) and Irwin et al (2005).

\begin{figure}[t]
\begin{center}
\includegraphics[width=\linewidth,clip]{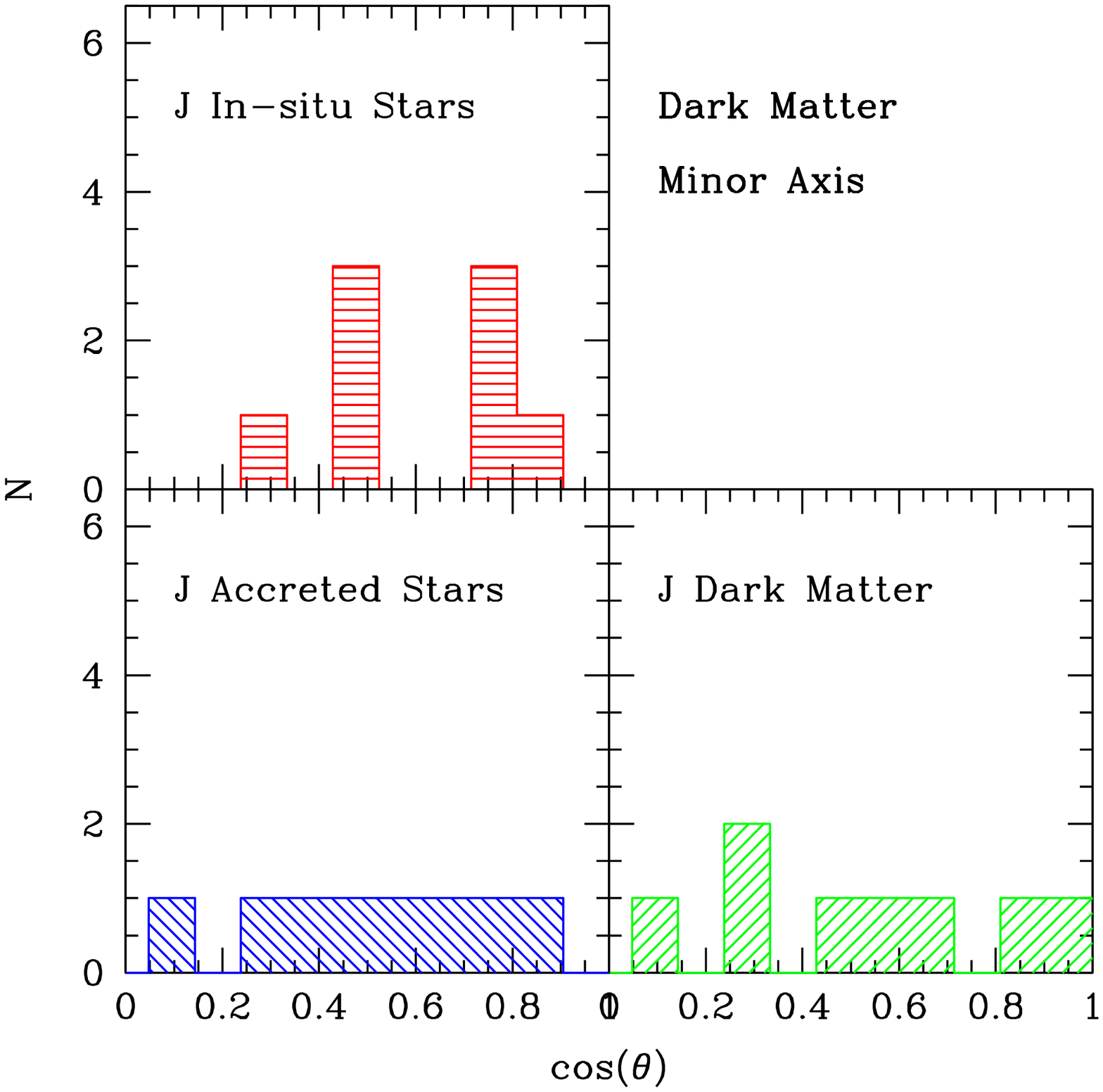}
\includegraphics[width=\linewidth,clip]{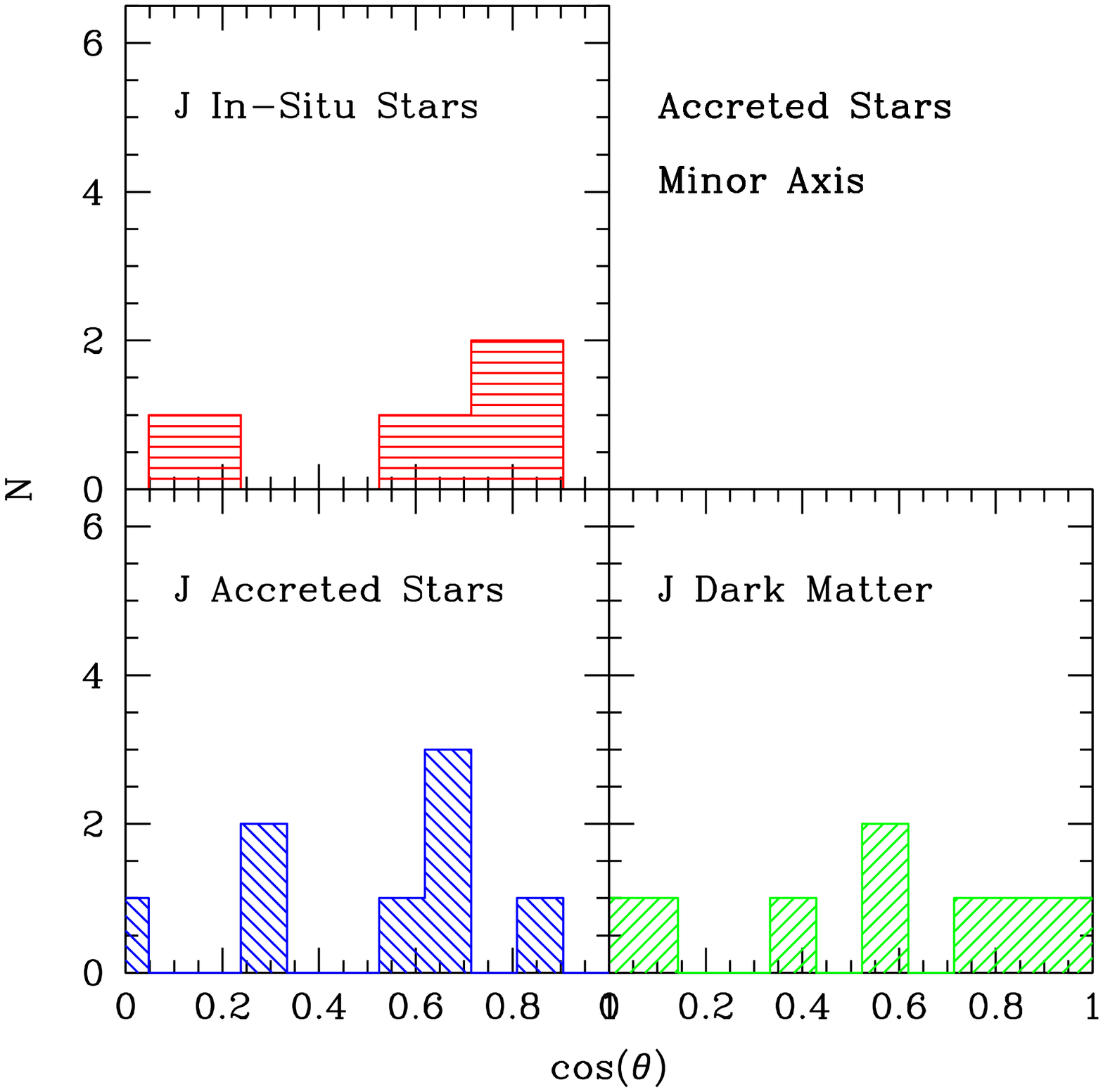}
\end{center}
\caption{ {\it (a)-top:} Distribution of the cosine of the angle between the
angular momentum of in-situ stars, ${\vec J}_{\rm ins}$, accreted stars, ${\bf
J}_{\rm acc}$, and the dark matter, ${\vec J}_{\rm drk}$, and the minor axis
of the dark matter . As in Figure~\ref{figs:jjdist}, histograms of different
shades refer to the various permutations, as indicated in the labels. {\it
(b)-bottom:} Same as (a), but for the minor axis of the accreted stellar
component.
\label{figs:jcdist}}
\end{figure} 

Quantitatively, we expect the slope of the surface brightness profile of the
outer luminous halo to steepen gradually outwards, as dictated by the
Sersic-law fits summarized in Table~\ref{tab:table2}. Given that the average
effective radii of the accreted component is $7.7$ kpc and $\langle n \rangle=
6.3$, we expect its surface brightness profile to steepen from from $\Sigma
\propto R^{-2.2}$ at the luminous radius ($20$ kpc) to $\Sigma \propto
R^{-2.9}$ at $r\sim 100$ kpc. This is in reasonable agreement with the results
of Zibetti et al (2004), Irwin et al (2005), and Guhathakurta et al
(2005). The latter authors, in particular, report that the M31 outer halo may
be approximated by $\Sigma \propto R^{-2.3}$ in the range $30$ kpc $<R< 100$
kpc, slightly shallower than in our simulations, but not inconsistent given
the substantial observational uncertainty.

\begin{figure}
\begin{center}
\includegraphics[width=\linewidth,clip]{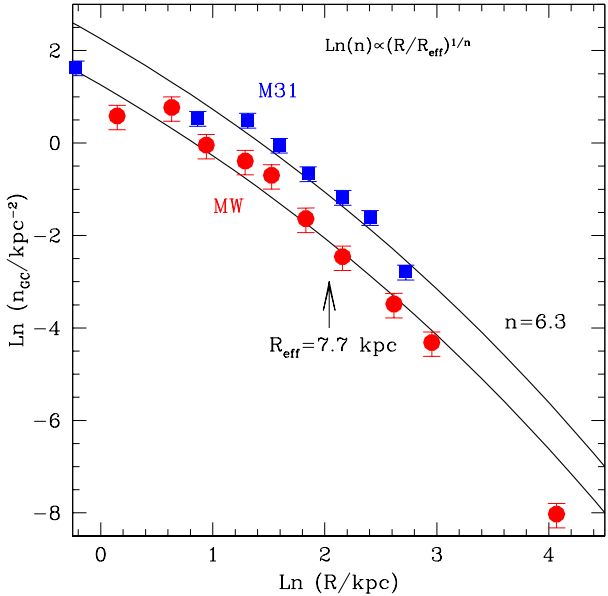}
\end{center}
\caption{ Projected number density profile of globular clusters around the
  Milky Way (filled circles) and M31 (filled squares), respectively. Data for
  the 147 Galactic globulars and for the 363 M31 globulars are from Harris
  (1996) and Galleti et al (2004). We assume that $100$ arcmin$=22$ kpc at the
  distance of M31. We have averaged the distribution of Galactic globulars
  over several random lines of sight to construct the projected number density
  profile. Each bin contains 15 and 40 globulars for the Galaxy and M31,
  respectively, and has Poisson error bars. The two curves are {\it not} fits
  to the data; rather, they illustrate the distribution of accreted stars in
  our simulations. Parameters for the two Sersic-law curves are chosen to be
  the average of the data presented in Table~\ref{tab:table2} and are
  vertically scaled to match the data for each galaxy.
\label{figs:gc}}
\end{figure}

We note as well that the distribution of globular clusters in the Milky Way
and M31 is consistent with that of the accreted stellar component in the
simulations. With the caveat of small number statistics (only $\sim 150$
globular clusters around the Galaxy and $\sim 360$ around M31 are currently
confirmed; see Harris 1996 and Galleti et al 2004), it is interesting that a
Sersic law with $n \sim 6$ is consistent with the (projected) number density
profile of Galactic globulars. This is shown in Figure~\ref{figs:gc}, where
the profile of M31 and Galactic globulars is compared with the Sersic-law that
describes the average accreted stellar component in our simulations. We
emphasize that the curves shown are {\it not} fits, and therefore the general
agreement is quite suggestive. A simple interpretation, of course, is that the
globular cluster population shares the same ``external'' origin as that of
accreted stars in our simulations, and that they are thus relicts of the
accretion history of each galaxy. (Guhathakurta et al (2005) note as well that
the smooth halo of M31 red giant stars they detect out to $\sim 150$ kpc
mimics the distribution of its globular cluster population.) Taken together,
the evidence linking globular clusters with accretion events, as envisioned in
the classic model of Searle \& Zinn (1978), seems quite strong.

In terms of space density, the slope of the accreted stellar component is
close to $\rho \propto r^{-3.1}$ near the edge of the luminous galaxy. This is
in reasonably good agreement with that inferred from tracers of the stellar
halo of the Milky Way in the vicinity of the Sun (see, e.g., Zinn 1985, Saha
1985, Morrison et al 2000). Our simulations imply that the halo profile should
steepen gradually with radius and become significantly steeper than $r^{-3}$
in the outer regions. It is interesting to explore the consequences of this
gradual steepening for the interpretation of dynamical data in the outskirts
of the Milky Way.

Battaglia et al (2005) report that the velocity dispersion of halo tracers
such as blue horizontal branch stars and red giants declines from $\sim 120$
km/s at the edge of the luminous disk to $\sim 50$ km/s at $\sim 120$
kpc. These authors argue that the data favour a dark matter halo model sharply
truncated in the outer regions over the extended mass distributions predicted
for CDM halos, such as the Navarro, Frenk \& White (1996, 1997, NFW)
profile. Neither halo model, however, is completely satisfactory. Extreme
radial velocity anisotropy ($\beta\sim 1$) is required to reconcile the data
with an NFW halo model although the opposite (i.e., a tangentially anisotropic
velocity distribution, $\beta \sim -0.5$) is required for the truncated halo
model.  The latter seems an unlikely result given that the outer regions of
systems assembled hierarchically are typically dominated by radial motions
(see, e.g., Hansen \& Moore 2004 and references therein).

\begin{figure}
\begin{center}
\includegraphics[width=\linewidth,clip]{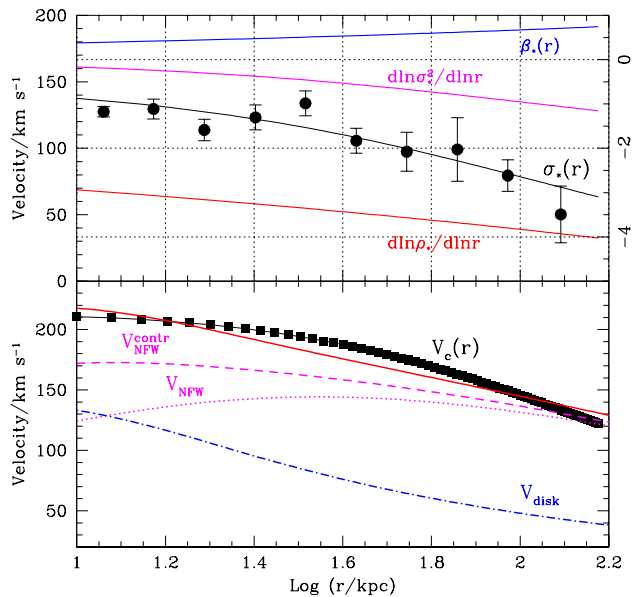}
\end{center}
\caption{{\it a-top:} Radial velocity dispersion of Milky Way stellar halo
  tracers (filled circles, from Battaglia et al 2005). These data are
  approximated with a simple model where $\sigma_{*}^2 \propto
  (1+r/r_0)^{-2}$, shown by the thin line through the data points. The three
  other curves (scale at right) illustrate the radial dependence of the
  right-hand side terms in Jeans' equation (eq.~\ref{eq:Jeans}), derived
  assuming a Sersic profile for the tracers. {\it b-bottom:} Solid squares
  show the {\it total} circular velocity of the Milky Way derived from the top
  panel. The top solid curve shows a fit assuming an (adiabatically
  contracted) NFW dark matter halo and an exponential disk. The contribution
  of the dark matter halo (before and after contraction), as well as that of
  the disk, are also shown. See text for further details.
\label{figs:vchalo}}
\end{figure}

We can use our results to explore the consistency of the velocity dispersion
data with CDM halos. We write the spherical Jeans' equation as,
\begin{equation}
V_c^2(r)=-\sigma_*^2 \bigl( {{\rm dln}\rho_* \over {\rm dln}r} +2\, \beta_*(r) +
{{\rm dln}\sigma^2_* \over {\rm dln}r} \bigr),
\label{eq:Jeans}
\end{equation}
where we have used a $*$ subscript to denote quantities associated with the
stellar halo. The terms in the left-hand side of eq~\ref{eq:Jeans} may be
evaluated by combining the Battaglia et al data with the assumption that the
stellar halo profile may be approximated by a Sersic-law with the (average)
parameters given in Table~\ref{tab:table2}. The result of this exercise is
shown in Figure~\ref{figs:vchalo}.

Interestingly, this simple model predicts a circular velocity near the center
consistent with that measured at the solar circle ($\sim 220$ km s$^{-1}$)
without tuning (solid squares in the bottom panel of
Figure~\ref{figs:vchalo}). Further out, the sustained drop in $\sigma_{*}$
leads to much lower circular velocities, approaching $\sim 150$ km s$^{-1}$ at
$r\sim 100$ kpc. This drop in the circular velocity of the system illustrates
why Battaglia et al's analysis favours a truncated halo. We argue, however,
that the data are consistent with an extended CDM halo, and that the velocity
dispersion drop places interesting constraints on its total mass. 

This is shown in the bottom panel of Figure~\ref{figs:vchalo} by the top thick
solid line, which corresponds to an (adiabatically contracted) NFW halo of
virial velocity $V_{\rm vir}=110$ km/s and concentration $c=14$ (the average
for halos of this mass in the $\Lambda$CDM cosmogony). The halo contribution
before and after contraction is shown by the dotted and dashed lines,
respectively.

The relatively good agreement with the $V_c$ profile derived from the data
implies that, rather than being truncated, the halo of the Milky Way may just
be less massive than commonly assumed. The virial mass of the fit shown in
Figure~\ref{figs:vchalo} is only $\sim 6.4 \times 10^{11} \, M_{\odot}$, and
its virial velocity is only one half of the circular speed at the solar
circle. This result is similar to that of Klypin, Zhao \& Somerville (2002),
who analyzed constraints placed by dynamical tracers within the luminous
radius of the Milky Way and found that reconciling such data with CDM halos
requires the virial velocity of the dark matter halo to be substantially lower
than the rotation speed of the disk. (Their favourite Milky Way halo model has
$M_{\rm vir} \approx 10^{12} \, M_{\odot}$, which corresponds to $r_{\rm
vir}\approx 260$ kpc, and $V_{\rm vir}\approx 130$ km s$^{-1}$.)  One may
rephrase this result by stating that the relatively small dark matter
contained within the luminous radius of the Galaxy is easiest to accommodate
with low-mass halo models, which have the virtue of predicting directly a
sharp drop in the velocity dispersion of outer halo tracers. A similar
argument was developed by Navarro \& Steinmetz (2000), although note the
erratum presented in Eke, Navarro \& Steinmetz (2001).

\begin{figure}
\begin{center}
\includegraphics[width=\linewidth,clip]{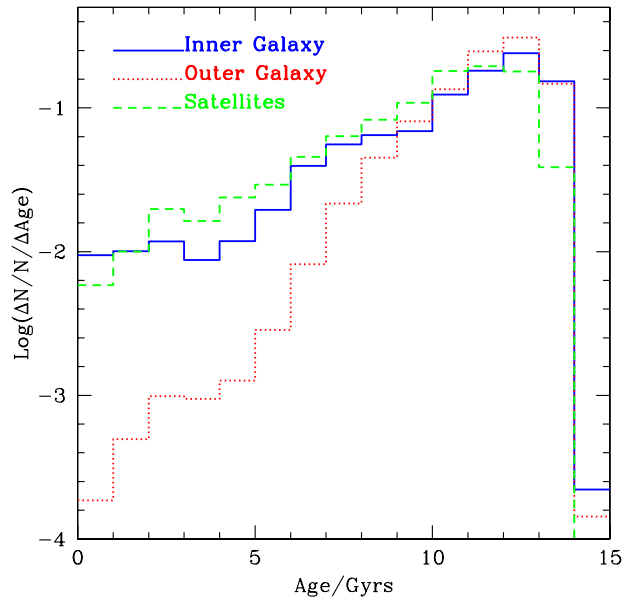}
\end{center}
\caption{ Age distribution of stars in the inner galaxy (i.e., stars within
  $r_{\rm lum}$, solid blue line), in the outer galaxy ($r>r_{\rm lum}$,
  dotted red line) and in satellites (dashed green line) enclosed inside the
  virial radius $r_{vir}$ and averaged over our set of 8 simulations.
\label{figs:agedist}}
\end{figure} 

It is unclear whether the mass of the halo of the Milky Way is unusually low
compared with other spiral galaxies of similar luminosity, but galaxy-galaxy
weak lensing data seem to favour halo masses of the order we find here for
late-type $L_{*}$ galaxies (see, e.g., Guzik \& Seljak 2002). Also, there is
now convincing evidence that the velocity dispersion of satellite galaxies
drops at large projected radii so the drop in $V_c$ in the outer regions may
actually be a common feature in Milky Way-like galaxies (see Brainerd 2004 for
a review).

Turning our attention now to the velocity structure of the outer halo, we
recall the pronounced radial anisotropy shown in Figure~\ref{figs:sigmar}, and
argue that such effect should not be ignored when interpreting the dynamics of
halo tracers in the outskirts of galaxies. As shown by Dekel et al (2005),
accounting for such anisotropies may be enough to reconcile the steeply
declining velocity dispersion profiles of planetary nebulae around bright
ellipticals (Romanowsky et al. 2003) with the presence of massive dark halos
around these objects. The radial anisotropy differentiates dynamically the
stars in the outer halo from the dark matter (Figure~\ref{figs:sigmar}). It
also distinguishes the stellar halo from the population of surviving
satellites, which are found to trace rather faithfully the main properties of
the dark halo. Indeed, the average anisotropy of the satellite population in
our simulations is found to be $\beta\sim 0.45$ (Sales et al, in preparation),
closer to that of the dark matter, and certainly less anisotropic than the
accreted stellar component.

This dynamical distinction is not surprising, as we find that the smooth
stellar halo has little relation with the population of surviving satellites
(see Figure~\ref{figs:agedist}). Indeed, few stars in the outer halo may be
traced to satellites that survive as self-bound entities until the
present. The properties of the stellar halo seems more closely linked to the
progenitors of early mergers than to the ``harassment'' of the satellite
population that survives until the present day. 

This result has interesting consequences when applied to the interpretation of
the origin of the intracluster light (ICL) component. Zibetti et al (2005),
for example, find that the properties of the ICL in their sample of clusters
selected from the SDSS are quite different from those of the cluster galaxy
population as a whole. In particular, the ICL is found to be more centrally
concentrated than the cluster galaxies and more significantly aligned with the
(brightest cluster galaxy (BCG) than cluster galaxies. In other words, the ICL
seems more intimately related to the central galaxy (BCG) than to the cluster
galaxy population.  All of these properties are consistent with a scenario
where the origin of the ICL is traced to the mergers that led to the formation
of the BCG, with a minor contribution from stars stripped from surviving
galaxies.

A final item for discussion is the weak alignment between dark halo shapes and
angular momentum shown in Figure~\ref{figs:jcdist}. At first, this seems in
disagreement with previous studies, many of which report significant alignment
between the rotation axis and the minor axis of CDM halos (see, e.g., Bailin
\& Steinmetz 2004 and references therein). The issue may be resolved by noting
that the shapes of the dark halos change dramatically when a dissipative
baryonic component is included; indeed, in most cases the dark halos are so
nearly spherical that the precise directions of the principal axes are poorly
determined. Furthermore, as discussed by Bailin et al (2005), it appears as if
the {\it inner} regions of the dark matter halo respond to the assembly of the
galaxy by aligning its minor axis with that of the luminous component, but at
the same time decouple from the orientation of the halo in the outer regions.

Interestingly, the weak correlations between halo shape and spin that our
simulations predict could be tested directly with upcoming weak lensing
surveys. Hoekstra et al (2004) report a tantalizing correlation between the
shape of the luminous galaxy and that of their surrounding matter, but their
sample contains a broad mixture of morphological types and a relatively strong
contribution from spheroid-dominated galaxies, where the relationship between
shape and spin is less clear than in spiral galaxies. Weak lensing analysis
applied to spiral-dominated samples should provide further constraints on this
topic. This is currently being addressed using large,
morphologically-segregated samples from the SDSS (Mandelbaum et al, in
preparation).

\section{Summary}
\label{sec:conc}

We investigate the origin of stars that populate regions well beyond the
traditional luminous boundaries of normal galaxies using numerical simulations
of galaxy formation in the $\Lambda$CDM cosmogony. Our simulations shows that
such components are ubiquitous and owe their presence to the many mergers that
characterize the early formation history of galaxies assembled
hierarchically. These faint stellar halos extend out to the virial radius of
the system, and consist mainly of the overlap and mixing of the many ``tidal
tails'' shed by merging progenitors during the assembly of the galaxy. Such
origin leads to robust predictions for the dynamics and structure of the outer
spheroid that may be contrasted with observation.

\begin{itemize}

\item The density profile of the accreted stellar component---which dominates
the outer luminous halo---is well approximated by a Sersic-like model where
the logarithmic slope steepens monotonically with radius; from $\rho \propto
r^{-3}$ near the edge of the galaxy's traditional luminous boundary to
$r^{-4}$ or steeper near the virial radius. The shape and concentration of the
accreted component is in reasonable agreement with the Galactic and M31
globular cluster population, lending support to the classic Searle-Zinn
scenario for the formation of the globular cluster population.

\item The accreted stellar component is reasonably well approximated with a
mildly triaxial spheroid with average axis ratios $\langle b/a \rangle \sim
0.91$ and $\langle c/a \rangle \sim 0.84$.  Rotation plays a negligible role
in the support of the accreted stellar component, which is characterized by a
strong radial anisotropy in its velocity distribution. This anisotropy grows
from the inside out, from $\sigma_r^2 \sim 2 \,\sigma_t^2$ in the luminous
outskirts of the galaxy to $\sigma_r^2 \sim 5 \,\sigma_t^2$ near the virial
radius.

\item The accreted stellar component is distinct from the dark matter halo as
well as from the satellite population, which are typically less concentrated
and more isotropic. Most stars in the outer halo formed in progenitors that
have since merged with the central galaxy. Only a small fraction of the outer
halo stars are contributed through ``harassment'' of satellites surviving as
self-bound entities until the present.

\end{itemize}

These properties are in broad agreement with recent observations of the
outskirts of spiral galaxies as well as of giant stars in the outer regions of
the Milky Way and of M31. In particular, they show that the outer light
``excess'' over extrapolations of the inner luminous profile recently reported
for M31 and other spirals is a generic feature of galaxies formed
hierarchically, and that the decline in velocity dispersion seen in the outer
Milky Way halo tracers is consistent with the presence of an extended, massive
halo of dark matter (albeit one of perhaps lower mass than commonly
assumed). They also imply that the intergalactic stellar component of galaxy
clusters may be more intimately related to the central galaxy than to the
galaxy cluster population as a whole. These results illustrate ways to unravel
the clues to the tumultuous accretion history of individual galaxies contained
in the stars ejected beyond their borders.

\section*{Acknowledgments}

We acknowledge useful discussions with Simon White, Stefano Zibetti, and
Vincent Eke. We thank Laura Sales for allowing us to quote some of her results
in advance of publication and Giuseppina Battaglia for providing the data used
in Figure~\ref{figs:vchalo} in electronic form. JFN acknowledges support from
NSERC, the Canadian Foundation for Innovation, as well as from the Alexander
von Humboldt and Leverhulme Foundations.

\vskip 5cm

\begin{table*}
\caption{Structural parameters of dark and luminous components of simulated
  galaxies}
\label{tab:table1}
\begin{tabular}{cccccccccc}
Label& $r_{\rm vir}$ & $M_{\rm vir}$ & $M_{\rm drk}$ & $M_{\rm str}$ & $M_{\rm
  glx}$ & $M_{\rm sat}$ & $M_{\rm out}$ & $f_{\rm glx}^{\rm acc}$ & $f_{\rm out}^{\rm acc}$\\
& kpc &  [$10^{11}M_{\odot}$] & [$10^{11}M_{\odot}$] & [$10^{11}M_{\odot}$] &
  [$10^{11}M_{\odot}$]& [$10^{11}M_{\odot}$]& [$10^{11}M_{\odot}$]&&\\\hline\hline
KIA1 & 391.53 &  29.47 &  25.96 &   3.32 &   2.36 &   0.47 &   0.48 &   0.38 &   0.91 \\
KIA2 & 266.41 &   9.28 &   7.88 &   1.29 &   1.04 &   0.15 &   0.10 &   0.30 &   0.95 \\
KIA3 & 267.57 &   9.41 &   7.92 &   1.25 &   1.02 &   0.10 &   0.12 &   0.42 &   0.96 \\
KIA4 & 350.05 &  21.06 &  17.96 &   2.75 &   1.91 &   0.51 &   0.33 &   0.56 &   0.94 \\
KIA5 & 316.30 &  17.61 &  15.13 &   2.25 &   1.68 &   0.35 &   0.23 &   0.40 &   0.96 \\
KIB1 & 394.00 &  31.53 &  27.19 &   3.53 &   2.60 &   0.55 &   0.37 &   0.39 &   0.93 \\
KIB2 & 269.59 &   9.62 &   8.18 &   1.33 &   1.04 &   0.17 &   0.11 &   0.38 &   0.97 \\
KIB3 & 267.83 &   9.43 &   7.94 &   1.23 &   0.97 &   0.13 &   0.12 &   0.49 &   0.95 \\
\end{tabular}
\vskip 0.5cm $M_{\rm drk}$ and $M_{\rm str}$ are, respectively, the dark and
stellar mass within the virial radius, $r_{\rm vir}$, defined to encompass a
region $100$ times denser than the critical density for closure. $M_{\rm glx}$
is the total mass of stars within the luminous radius, $r_{\rm lum}=20$
kpc. $M_{\rm sat}$ is the mass of stars in satellites. $M_{\rm out}$ is the
mass of stars outside the luminous radius but not in satellites. $f_{\rm
glx}^{\rm acc}$ and $f_{\rm out}^{\rm acc}$ are the fractions of accreted
stars in the inner ($r<r_{\rm lum}$) and outer ($r>r_{\rm lum}$) galaxy,
respectively.
\end{table*}

\begin{table*}
\centering
\caption{Parameters of Sersic-law fits to $R$-band surface brightness profile
of {\it accreted} stars in the outer ($30<R/{\rm kpc}<130$) galaxy.}
\label{tab:table2}
\begin{tabular}{cccccc}
Label & $n$ & $R_{\rm eff}$ & $\mu_{\rm eff}$ & $L_{\rm tot}^{\rm fit}$ &
$L_{\rm tot}^{\rm acc}$\\
& & kpc & mag/asec$^2$ & [$10^{10}L_{\odot}$] & [$10^{10}L_{\odot}$] \\\hline\hline
KIA1 & 5.01 & 12.41 & 23.62  &  3.15 &  3.68 \\ 
KIA2 & 6.13 &  6.05 & 23.48  &  0.85 &  1.03 \\
KIA3 & 5.73 &  5.45 & 22.93  &  1.10 &  1.44 \\
KIA4 & 5.99 &  6.33 & 22.36  &  2.61 &  3.52 \\
KIA5 & 8.27 & 11.06 & 24.54  &  1.23 &  2.07 \\
KIB1 & 6.95 &  8.49 & 23.10  &  2.57 &  3.78 \\
KIB2 & 6.10 &  6.07 & 23.30  &  1.01 &  1.32 \\
KIB3 & 5.96 &  5.62 & 23.15  &  0.97 &  1.90 \\
\end{tabular}
\vskip 0.5cm 
$L_{\rm tot}^{\rm fit}$ is the total luminosity of the Sersic fit
to the outer profile. $L_{\rm tot}^{\rm acc}$ is the total luminosity of the
accreted stellar component.
\end{table*}

\end{document}